\documentclass[%
 reprint,
 superscriptaddress,
 amsmath,amssymb,
 aps,
 prl,
]{revtex4-1}

\usepackage{graphicx}
\usepackage{dcolumn}
\usepackage{bm}
\usepackage{hyperref}

\hypersetup{
	colorlinks = true,
	pdftitle = {},
	pdfauthor = {},
	pdfkeywords = {},
	linkcolor = blue,
	citecolor = blue,
	filecolor = black,
	urlcolor = magenta
}

\begin{document}

\title{Proposal for measuring magnetism with patterned apertures}%

\author{Devendra Negi}
\author{Jakob Spiegelberg}
\affiliation{Department of Physics and Astronomy, Uppsala University, P.O. Box 516, 75120 Uppsala, Sweden}%
\author{Shunsuke Muto}
\affiliation{Electron Nanoscopy Section, Advanced Measurement Technology Center, Institute of Materials and Systems for Sustainability, Nagoya University, Furo-cho, Chikusa-ku, Nagoya 464-8603, Japan}
\author{Thomas Thersleff}
\affiliation{Department of Materials and Environmental Chemistry, Stockholm University, Svante Arrhenius v\"{a}g 16C, 106 91 Stockholm, Sweden}
\author{Masahiro Ohtsuka}
\affiliation{Department of Materials Physics, Graduate School of Engineering, Nagoya University, Furo-cho, Chikusa-ku, Nagoya 464-8603, Japan}
\author{Linus Sch\"{o}nstr\"{o}m}
\affiliation{Department of Materials and Environmental Chemistry, Stockholm University, Svante Arrhenius v\"{a}g 16C, 106 91 Stockholm, Sweden}
\author{Kazuyoshi Tatsumi}
\affiliation{Advanced Measurement Technology Center, Institute of Materials and Systems for Sustainability, Nagoya University, Chikusa, Nagoya 464-8603, Japan}
\author{J\'{a}n Rusz}%
\email{jan.rusz@physics.uu.se}
\affiliation{Department of Physics and Astronomy, Uppsala University, P.O. Box 516, 75120 Uppsala, Sweden}%

\date{\today}

\begin{abstract}
We propose a magnetic measurement method utilizing a patterned post-sample aperture in a transmission electron microscope. While utilizing electron magnetic circular dichroism, the method circumvents previous needs to shape the electron probe to an electron vortex beam or astigmatic beam. The method can be implemented in standard scanning transmission electron microscopes by replacing the {spectrometer} entrance aperture with a specially shaped aperture, hereafter called \emph{ventilator aperture}. The proposed setup is expected to work across the whole range of beam sizes -- from wide parallel beams down to atomic resolution magnetic spectrum imaging.
\end{abstract}


\maketitle

Nanotechnologies utilizing magnetic materials call for characterization techniques that allow to quantify magnetic properties with sufficient spatial resolution. Typically used methods, such as spin-polarized scanning tunneling microscopy \cite{spstm,spstmafm}, magnetic exchange force microscopy \cite{magefm}, x-ray magnetic circular dichroism \cite{xmcd15,xmcdptycho} or electron holography \cite{magholo}, are either restricted to surface analyses or they lack sufficient spatial resolution.

An alternative measurement technique called electron magnetic circular dichroism (EMCD) is under development since its proposal in 2003 \cite{emcdproposal} and experimental confirmation in 2006 \cite{nature}. EMCD utilizes (scanning) transmission electron microscopes [(S)TEM], therefore it is a natural candidate to go beyond the previous limitations aiming for atomic spatial resolution magnetic measurements. Since 2006, significant improvements have been achieved in this direction, recently measuring antiferromagnets with astigmatic beams of atomic size \cite{c34exp}, or utilizing aberration-free electron probes in experimental geometries with suitably oriented crystalline samples \cite{apremcd,largealpha}. Most recently, an alternative setup based on high-resolution TEM imaging has allowed for the detection of quantitative magnetic information from individual atomic planes \cite{elspemcd}.

{EMCD measurements based on the initial experiment \cite{nature} use an off-axis aperture, most commonly a circular one built into the spectrometer, see e.g.\ \cite{emcd2nm,salafranca,dongshengapl,largealpha,fudyfe2,sebastianfept,tomfe3o4,tominterface} among others. While many times successful, this approach has several shortcoming. Circular aperture collects only a small fraction of the inelastically scattered electrons and it is not an optimal shape for EMCD acquisition \cite{verbeeckaperture}. Its usual implementation then involves tilting the sample into two-beam or three-beam orientation, losing the view of individual atomic columns. In a STEM implementation it requires acquisition of several spectrum images from the same sample area \cite{emcd2nm,salafranca,tomfe3o4,largealpha,tominterface}. This becomes progressively more challenging, when the desired spatial resolution increases together with convergence angles \cite{lofflerconv}. Alternative ways of detection were followed in Refs.~\cite{schattsuperstem,schachinger}, though in these approaches it was not yet possible to extract EMCD spectra due to low signal to noise ratios.}

{
In this Letter, we describe three crucial findings that have arisen from simulations of magnetically-sensitive inelastic scattering.  First, we propose a new approach to optimize the EMCD signal collection in a wide range of crystallographic symmetries and scattering geometries.  Using cubic systems as a test bed, we demonstrate that a custom-designed aperture can selectively allow for an optimal collection of either the positive or negative EMCD signal in a variety of high-symmetry zone axes.  Second, we show that the signal collection with these apertures is robust across a large range of convergence angles, including those sufficiently large to focus an electron probe down to less than the width of an atomic column.  Finally, we propose a strategy to allow for the collection of both positive and negative EMCD signal contributions simultaneously, negating the need for scanning the same sample area multiple times.  These three findings represent a route to overcome the long-standing issues with EMCD acquisition.
}


{Before discussing possible paths of practical realization of the scheme, we employ simulations to motivate our approach.} The angle-resolved inelastic electron scattering cross-section is simulated using the combined multislice / Bloch waves method \cite{vortexsurvey} as implemented in the code \textsc{mats.v2} \cite{matsv2}. Simulations are performed at an acceleration voltage set to 200~kV for a crystal of bcc iron in two orientations: [001] zone axis and a three-beam orientation with the systematic row of reflections parallel to $\mathbf{G}=(110)$ \cite{emcd3bcdeloc}.

\begin{figure}
  \includegraphics[width=8.6cm]{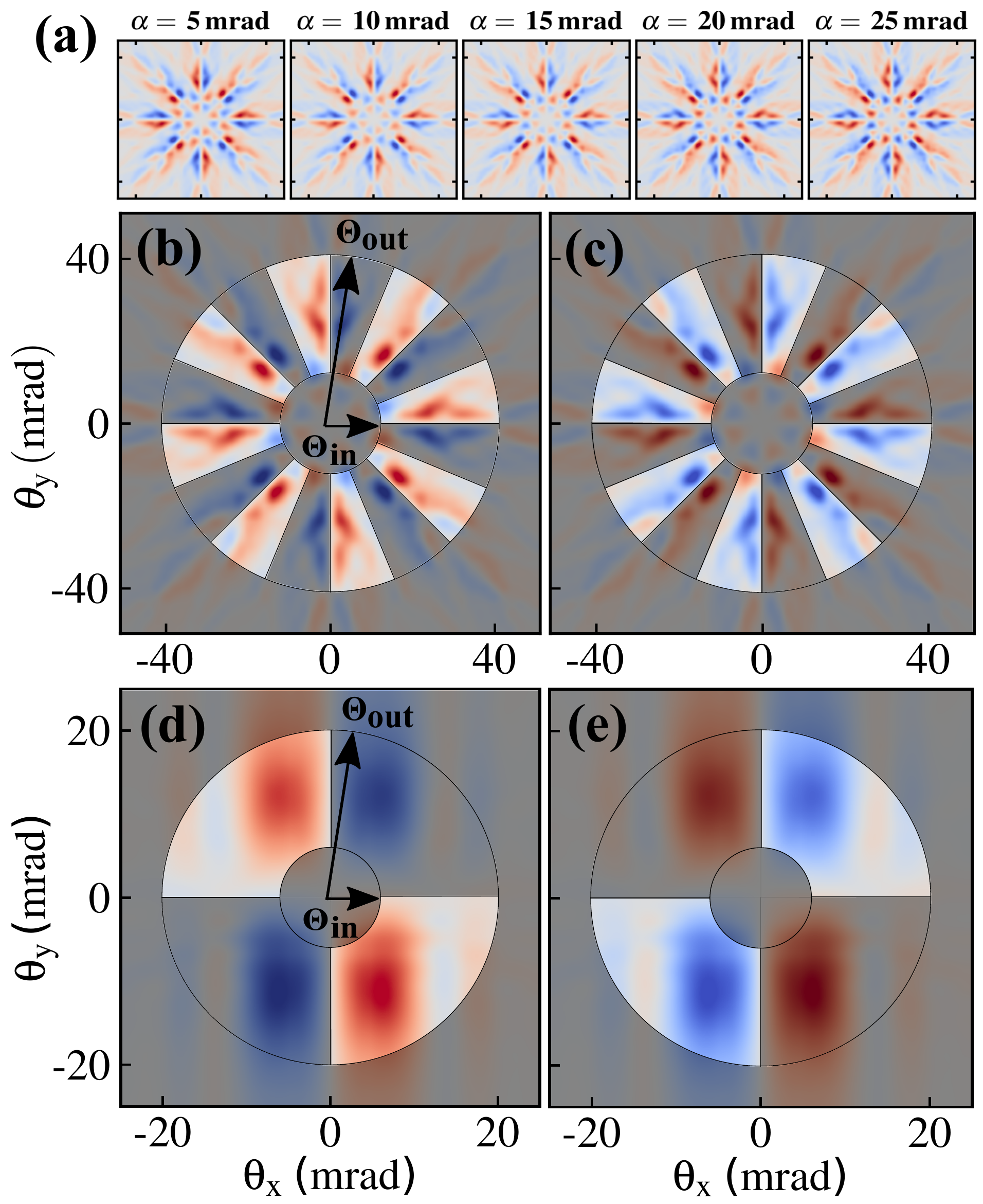}
  \caption{a) Examples of energy filtered $L_3$ edge diffraction patterns of bcc Fe in [001] zone axis orientation calculated {for sample thickness $t=18$~nm} at 200~kV acceleration voltage showing distribution of the magnetic signal in the diffraction plane for various convergence {semi-angles $\alpha$}. In b) and c) a ventilator aperture is applied in its two orientations, letting pass through mostly positive or negative EMCD signal, respectively {($\alpha=10$~mrad, $t=18$~nm)}. In d) and e) the same principle is applied to a 3-beam geometry, for a ventilator aperture with two blades {($\alpha=10$~mrad, $t=19$~nm)}.}
  \label{fig:scheme}
\end{figure}

Simulations and measurements of the EMCD distribution in the diffraction plane in the three-beam orientation have been reported in the literature \cite{emcd2nm,hansprl,opmaps,asym3bc,feptnp,apremcd,largealpha,inplane,lofflerconv}. The general qualitative distribution of the magnetic signal is that the EMCD signal has a dominant sign in a given quadrant of the diffraction plane, which changes in the neighboring quadrants. This led to the introduction of the double-difference procedure \cite{hansprl,feptnp,largealpha} utilizing circular {aperture to acquire electron energy-loss spectra one-by-one from each quadrant sequentially}.

Fewer EMCD simulations have been done in a zone axis orientation \cite{opmaps,zasong}, nevertheless they also show certain universality in the distribution of EMCD in the diffraction pattern, the magnetic signal being antisymmetric with respect to all the symmetry axes (horizontal, vertical and both diagonal ones). {In addition, with respect to the center of the diffraction pattern, the EMCD signal has a rotational symmetry whose order, however, depends on the crystal at hand.}

These universalities lead us to propose a patterned detector aperture, which would have a regular rotationally symmetric shape dictated by rotational symmetry of the crystal and its orientation. In the case of [001] zone axis orientation of bcc Fe, it would be symmetric with respect to the rotations by 90 degrees, while in the three-beam orientation it is symmetric with respect to a rotation by 180 degrees. This leads to a characteristic ventilator-like shape, thus the name \emph{ventilator aperture}. 

Figure~\ref{fig:scheme} shows a scheme of the principle. In Fig.~\ref{fig:scheme}(a) we present a set of simulations showing magnetic signal distribution in the [001] zone axis Fe-$L_3$ diffraction pattern for various convergence semi-angles $\alpha$. The universal pattern of EMCD distribution can be recognized. In Fig.~\ref{fig:scheme}(b),(c) and Fig.~\ref{fig:scheme}(d),(e) we then show the ventilator apertures overlaid on the diffraction patterns in a zone axis and 3-beam orientation, respectively. Note how a ventilator shaped aperture permits selection of predominantly positive (red) or negative (blue) regions of the EMCD distribution {by rotating 22.5 degrees and 90 degrees for zone axis and 3-beam orientation, respectively. (Suitable number of blades and the rotation angle obviously depend on the crystal symmetry and its orientation with respect to the electron beam, e.g., hexagonal crystals would require 3- or 6-blade apertures, etc.)
}

\begin{figure}
  \includegraphics[width=8.4cm]{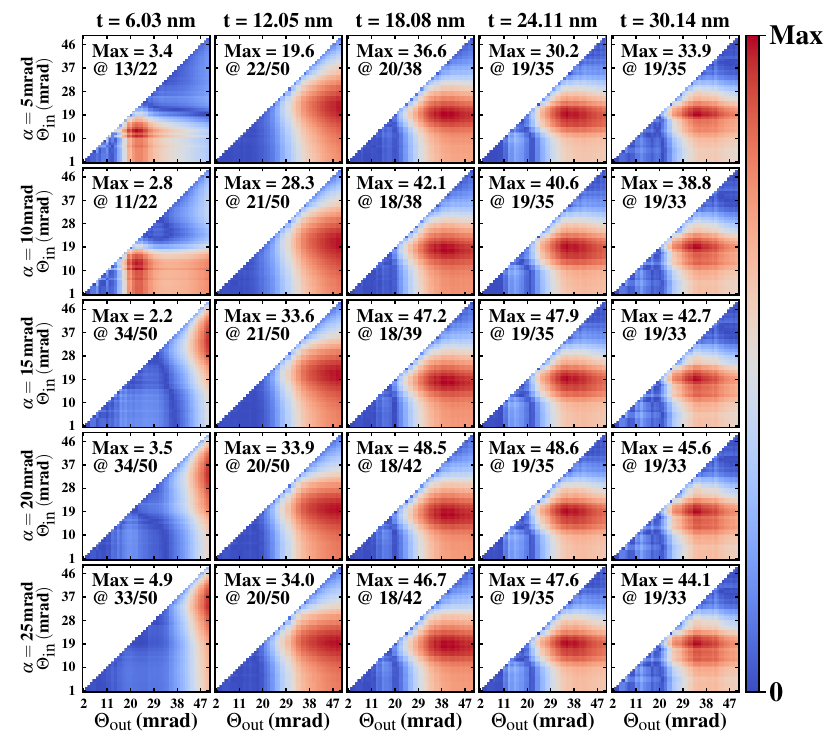}
  \includegraphics[width=8.4cm]{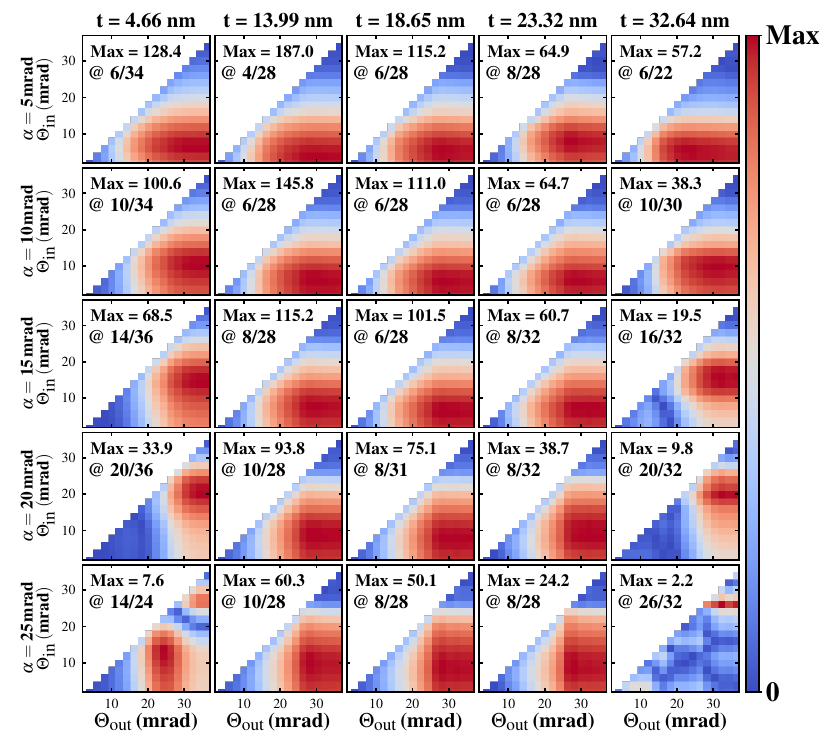}
  \caption{Optimization of the signal to noise ratio (see Eq.~\ref{eq:snr}) for a) zone axis and b) 3-beam orientations, respectively, as a function of inner and outer collection angles, $\Theta_\text{in}$ and $\Theta_\text{out}$, for a range of sample thicknesses {$t$} and convergence {semi-angles $\alpha$. Each panel shows the maximal SNR (in arbitrary units) as well as the inner/outer collection semi-angles (in miliradians) for which it was reached.}  All values are normalized to the same electron dose {indicent on the sample} and the electron beam is always centered on an atomic column.}
  \label{fig:snropt}
\end{figure}

While the symmetry (number of blades) of the ventilator aperture is fixed by the symmetry of the crystal and its orientation with respect to the electron beam, the inner and outer collection angles, $\Theta_\text{in}$ and $\Theta_\text{out}$, are free parameters. A natural choice of these parameters would be such that maximizes the signal-to-noise ratio (SNR) of the EMCD signal, which can be expressed as \cite{c34opt}
\begin{equation} \label{eq:snr}
  SNR = f_{red} \frac{M}{N} \frac{\sigma_{mag}}{\sigma_{nm}} \sqrt{\frac{2C_{L_3}}{1+b}}
\end{equation}
where $M/N$ is a material dependent property, $C_{L_3}$ is count of electrons detected within the $L_3$-edge energy range after the background subtraction and $b$ is a ratio of the background electron counts to $C_{L_3}$ within the same energy range. Finally, $\sigma_{mag}/\sigma_{nm}$ is a ratio of normalized scattering cross-sections computed with mixed dynamical form-factor \cite{kohl} set to $S(\mathbf{q,q'})=i(q_x q'_y - q'_x q_y)$ and $S(\mathbf{q,q'})=\mathbf{q.q'}$, representing EMCD due to magnetization parallel to $z$-axis and the non-magnetic component of the scattering cross-section, respectively \cite{c34opt}.


Explicit optimization of the SNR for both zone-axis and 3-beam orientations is plotted in Fig.~\ref{fig:snropt}. Except for the thinnest samples considered in zone-axis orientation, the optimization over $\Theta_\text{in}$ and $\Theta_\text{out}$ leads to a very similar pattern, only weakly dependent on sample thickness {$t$} or convergence semi-angle {$\alpha$}. This unforeseen finding suggests that for each of the two symmetries (8-blade and 2-blade ventilator apertures, respectively) it should be possible to construct a universal ventilator aperture, which will be close-to-optimal for all convergence semi-angles and sample thicknesses. {(Note though that its inner and outer collection angles depend on material and its Bragg scattering angles, nevertheless.)} A global optimization is obviously an ambiguous procedure, due to the given (arbitrary) choice of convergence angles and thicknesses in our simulations, and also the weighting of the individual cases. Here we simply averaged all the SNRs over the panels presented in Fig.~\ref{fig:snropt}, 
from which we extracted the proposed optimal $\Theta_\text{in}=19$~mrad and $\Theta_\text{out}=38$~mrad for the zone axis case, and $\Theta_\text{in}=6$~mrad and $\Theta_\text{out}=30$~mrad for the 3-beam geometry.

We stress again that the exact values have little of meaning due to arbitrariness of the global optimization procedure. Nevertheless, some semi-quantitative observations can be made: 1) in the zone axis orientation the optimal collection of signal should happen at larger scattering angles, with $\Theta_\text{out} \approx 2\Theta_\text{in}$, 2) in 3-beam orientation the crystal scatters less strongly and consequently the $\Theta_\text{in}$ can be much smaller. On the other hand, $\Theta_\text{out}$ can be relatively large, therefore a much larger fraction of inelastically scattered electrons can be utilized for the EMCD measurements. This is reinforced by explicit comparison of the SNR values in Fig.~\ref{fig:snropt} between zone-axis condition and 3-beam geometry, which suggests that for a fixed electron dose the 3-beam orientation often offers a significantly better SNR than the zone-axis orientation, except for thicker samples measured with larger convergence angles. 

\begin{figure}
  \centering
  \includegraphics[width=8.6cm]{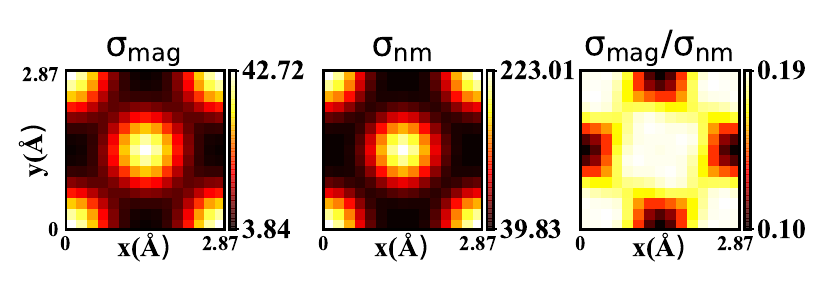}
  \includegraphics[width=8cm]{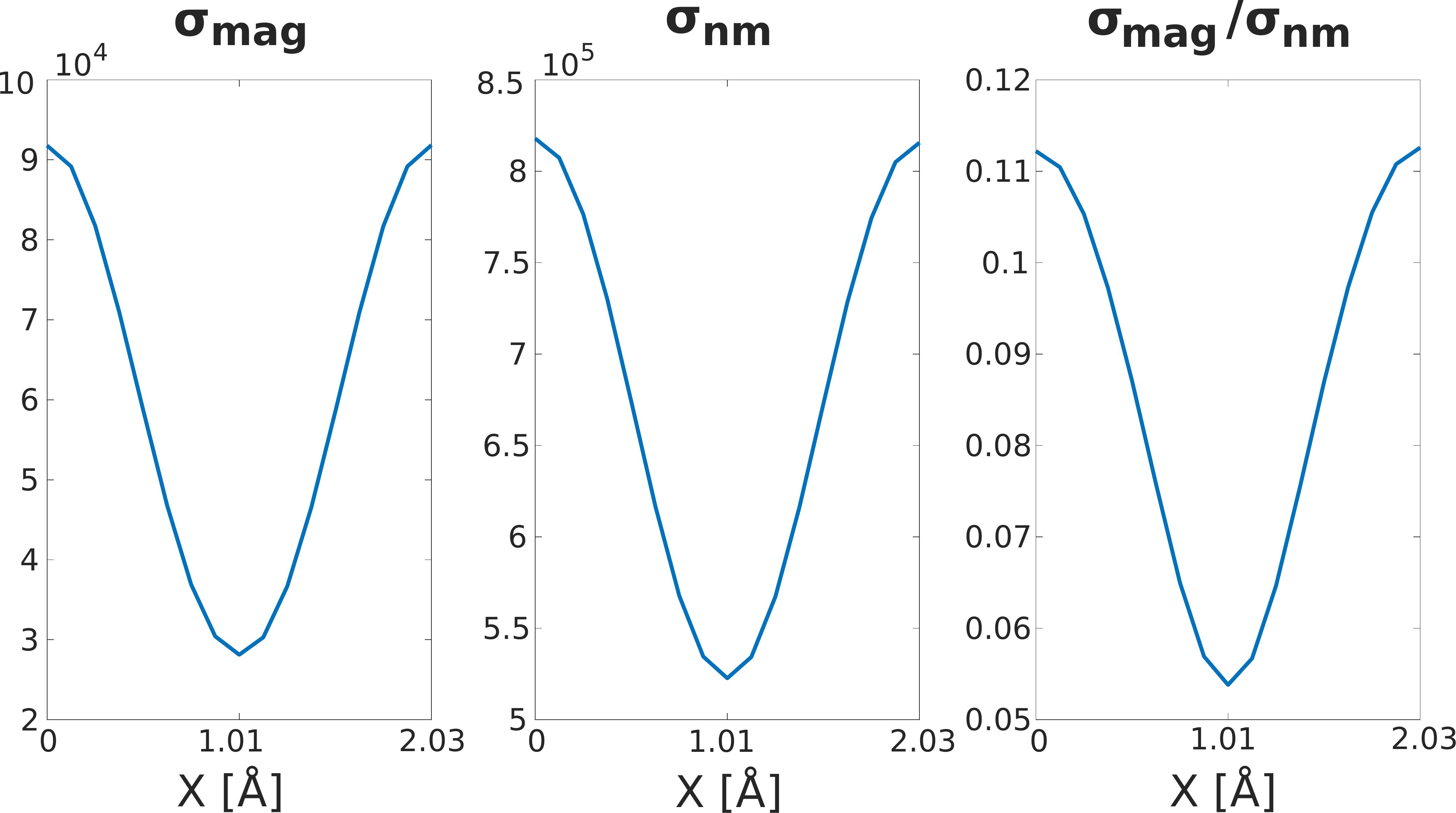}
  \caption{a) Spectrum image simulation for the [001] zone axis orientation of {18~nm thick} bcc Fe at 15~mrad convergence semi-angle and 200~kV acceleration voltage. Left, middle and right panels show maps of the magnetic signal, the non-magnetic signal and their ratio, respectively, as a function of beam position over a bcc iron unit cell. b) Linear profiles of magnetic signal, nonmgagnetic signal and their ratio, as a function of beam position in between atomic planes in 3-beam geometry {($\alpha=15$~mrad, $t=19$~nm)}. The distance between atomic planes is $d = a\frac{\sqrt{2}}{2} = 2.03$~\AA{}. Note that in both cases the magnetic signal stays non-negligible and positive.}
  \label{fig:stemsi}
\end{figure}

\begin{figure*}[t]
  \includegraphics[width=16cm]{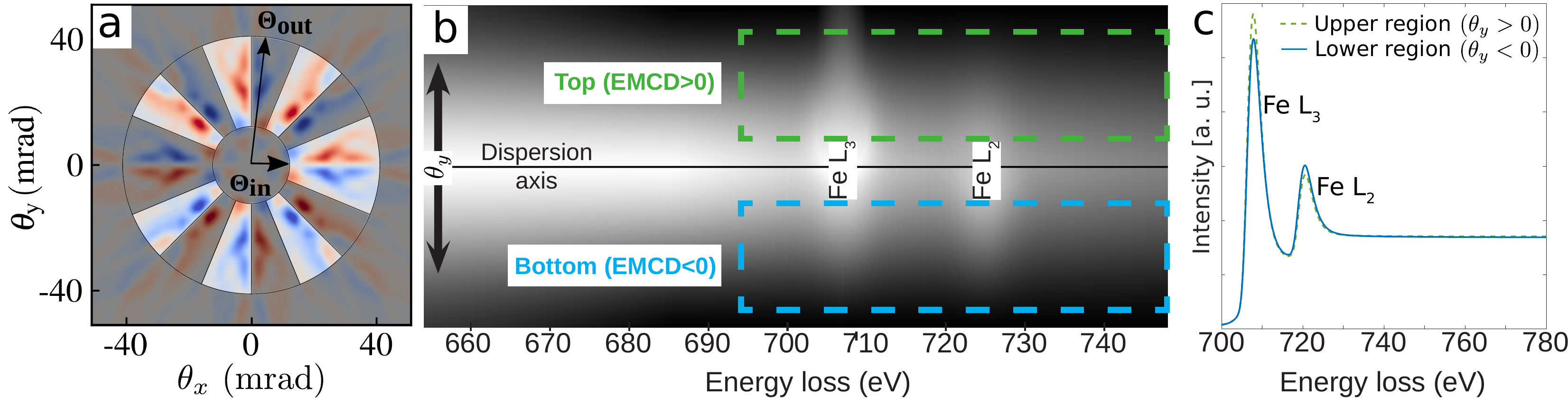}
  \caption{a) Mirrored ventilator aperture overlaid on a simulated distribution of the EMCD signal in the diffraction plane (see Fig.~\ref{fig:scheme}). b) Schematic image detected by CCD with its $x$-axis being an energy dispersion axis and $y$-axis aligned with $\theta_y$-axis of the aperture. c) Expected spectra extracted from the rectangular regions in panel b) leading to an EMCD signal as their difference.\label{fig:mirrorvent}}
\end{figure*}

Figure~\ref{fig:snropt} suggests that high SNRs can be obtained at rather large convergence semi-angles. Therefore, a natural question is, whether ventilator apertures would allow to measure magnetic properties with atomic resolution. Figure~\ref{fig:stemsi}(a) shows a simulation of an atomic resolution STEM spectrum image integrated over the Fe-$L_3$ edge for $\alpha=15$~mrad and $t=18$~nm. We show $\sigma_{mag}$, $\sigma_{nm}$ and their ratio separately. Note that atomic columns are clearly resolved both in magnetic and non-magnetic component of the scattering cross-section. Interestingly, the magnetic signal doesn't change sign throughout the whole unit cell, which is different from atomic resolution spectrum imaging calculated with vortex beams in Ref.~\cite{vortexelnes,vortexsurvey}. This is an important advantage allowing to use all the spectra in STEM spectrum image dataset for extraction of magnetic properties, when contrasting two sets with EMCD contributions of opposing signs (as measured with correspondingly rotated apertures).

{This observation deserves a more detailed discussion. Qualitatively, the reason is due to a different dominant mechanism of generating EMCD intensity by vortex beams, compared to classical EMCD approaches with convergent probes. Because of the symmetric on-axis placement of the detector entrance aperture, vortex beams allow to detect EMCD thanks to their favorable phase distribution in the central CBED disk, as is analyzed in detail in Ref.~\cite{emcdc34} -- the term $2\mathrm{Re}[e^{-i\Delta\phi_{\mathbf{k,G}}}S(\mathbf{q,q-G},E)]$ in Eq.~2, see Ref.~\cite{emcdc34} for details about the notation. The favorable phase distribution manifested in $\Delta\phi_{\mathbf{k,G}}$ gets distorted by beam shifts, which introduce an additional phase modification (a phase ramp). Classical EMCD with convergent beams, which uses off-axis detector entrance apertures, detects EMCD primarily due to the term $2T_\mathbf{G}\mathrm{Im}[S(\mathbf{q,q-G},E)]$, which is not sensitive to modifications of the phase distribution due to beam shifts. That is because beam shift in the first approximation modifies the phase distribution in the central CBED disk and Bragg scattered disks in the same way.}

STEM spectrum imaging with atomic resolution in the 3-beam orientation would show parallel stripes indicating positions of atomic planes and the only contrast appears in direction perpendicular to the planes \cite{apremcd}. Such linear profiles are plotted in Fig.~\ref{fig:stemsi}(b) for $\alpha=15$~mrad and $t=19$~nm. Similarly as in the zone axis orientation, the magnetic signal doesn't change sign throughout the scan, yet it shows sufficient contrast allowing identification of individual atomic planes.

{Now we turn our attention to paths to experimental realization of the proposal.
Probably the most convenient solution would be to modify the existing spectrometer entrance apertures and develop a mechanism for rotating them. Alternatively a custom aperture system could be placed at any point between sample and spectrometer by setting up the post-specimen optics in such a way that a diffraction pattern forms in the aperture plane.}

{We note that to obtain data allowing to extract EMCD spectra, it still requires acquisition of two spectrum images, one for each orientation of the ventilator aperture.} This requires both a physical rotation of either the aperture or the sample as well as illumination of the exact same sample area for both datasets. Fulfilling these requirements with atomic column resolution could be demanding.

To mitigate these challenges, we propose breaking the rotational symmetry of the aperture by instead exploiting the mirror symmetry about the dispersion axis, as is illustrated in Fig.~\ref{fig:mirrorvent}. Provided the full 2D CCD is recorded and that $\theta_x$ is aligned parallel to the dispersion axis of the spectrometer, such a geometry allows to extract spectra with opposite signs of EMCD from the two half-planes of the CCD. While similar to the Large Angle Convergent DIFfraction (LACDIF) geometry previously used to acquire an EMCD signal \cite{lacbed}, our proposed geometry allows the probe to be fully converged on the sample. This experimental design thus permits simultaneous acquisition of both signs of EMCD with a single scan while acting on the zone axis of a magnetic material with a fully converged probe under electron optical conditions suitable for atomic column resolution. {This strategy could be readily implemented by exchanging the spectrometer entrance aperture by a suitable patterned one, provided that we can align the energy dispersion axis parallel to $\theta_x$. Naturally, an (in this approach optional) physical rotation mechanism could simplify the alignment.}

{We have proposed a strategy for EMCD acquisition that should result in the ability to probe individual atomic columns.  First, we use simulations to derive a series of optimized patterned aperture designs that can be implemented in any desired zone axis geometry.  Second, we demonstrate that the EMCD signal is largely robust with respect to convergence angle, allowing for this strategy to be implemented for a wide variety of probe sizes ranging from above 10 nm down to sub-\AA{}ngstr\"{o}m in diameter.  Finally, we propose an aperture and spectrometer configuration that would allow for the simultaneous collection of the negative and positive EMCD contributions, negating the need for multiple scans.  
}


\begin{acknowledgments}
We acknowledge Klaus Leifer for helpful discussions. DSN, JS and JR acknowledge funding from the Swedish Research Council, G\"{o}ran Gustafsson's foundation, Carl Tryggers foundation and Center of Interdisciplinary Mathematics at Uppsala University. {SM and MO acknowledge Grants-in-Aid for Scientific Research on Innovative Areas ``Nano Informatics'' (No. 25106004) from the Japan Society of the Promotion of Science. TT and LS acknowledge funding from the Swedish Research Council (project 2016-05113) and the Knut and Alice Wallenberg Foundation (3DEM-NATUR 2012.0112).} Simulations have been performed using resources of the Swedish National Infrastructure for Computing (SNIC) at the NSC Center (cluster Triolith) and HPC2N Center (cluster Abisko).
\end{acknowledgments}

\end{document}